# A CLASS OF SADDLE-POINT CONFIGURATIONS IN THREE-DIMENSIONAL SU(2) LATTICE GAUGE THEORY

ROBERT D. MAWHINNEY*

*Department of Physics, Columbia University*
*New York, NY, 10027, USA*
E-mail: rdm@phys.columbia.edu

ABSTRACT

We discuss a class of saddle-point configurations in SU(2) lattice gauge theory in three Euclidean dimensions. These configurations are smooth on the scale of the lattice and have an action density exhibiting localized peaks, as has been seen in cooled and extremized Monte Carlo generated lattices. Large Wilson loops centered on the action peaks show a unit of Z(2) flux. We discuss the generation of these configurations and measurements of the Creutz ratios on them.

## 1. Introduction

A detailed fundamental description of the origins of confinement in non-abelian gauge theories has still not been found. We have extensive evidence from numerical simulations that non-abelian theories do confine at weak coupling, but distilling the essential properties of numerically generated configurations which produce this effect has not been accomplished. It is certainly a possibility that no reasonably simple way of understanding confinement exists, but this note concentrates on whether some understanding of the confinement mechanism can be gleaned from numerical simulations.

Cooling[1,2] (a deterministic reduction of the action) and extremization[3] (the evolution of a lattice towards the nearest, possibly unstable, solution of the lattice field equations) are two procedures which have been used to probe the features of Monte Carlo generated lattices. Both of these procedures, when applied to SU(2) lattices in three dimensions, produce configurations with isolated peaks in the action density. The quasi-stability of the peaks under cooling is due to their proximity to a solution (with presumably only a few unstable modes) of the field equations. (There are non-trivial solutions since the periodic lattice boundary conditions we use do not allow the scaling arguments needed in Derrick's theorem.)

One major difference between cooling and extremizing Monte Carlo generated lattices is the behavior of the string tension; it persists under moderate amounts of cooling and is not apparent in the extremized lattices. This is in spite of the similarity of the lumps produced with both procedures. The cooled lumps are imprecisely defined (they vanish with enough cooling), making the extremized lattices a better

*Contribution to the proceedings of Confinement '95. This work was done in collaboration with Chulwoo Jung. It was partially supported by the US Department of Energy and the Pittsburgh Supercomputing Center.

starting point for quantitative investigations. The loss of the string tension is a serious issue; however, if saddle-point configurations are important in evaluating the functional integral, they must be included with the proper weighting. The saddle points produced by extremizing Monte Carlo generated lattices may not occur with the correct weight, due to the detailed way in which the algorithm finds a saddle and the influence of high frequency modes on the saddles that are found.

To study the role of saddle-point configurations in SU(2)$_3$, we have used one property of the cooled/extremized lumps to allow us to make saddle-point configurations from known starting configurations. This is described in Section 2. Section 3 discusses the monopole content of one of our saddles in maximal abelian gauge and section 4 describes the results of measuring the string tension on individual saddle-point configurations.

## 2. Making Saddle-point Configurations

We want to study saddle-point configurations that display isolated peaks in the action density. The configurations must be smooth on the scale of the lattice to insure that we are not studying physics related to the underlying lattice. (This does not insure that we are studying continuum physics.) We also want the saddles we study to display as many of the properties of the cooled/extremized lumps as possible.

The starting point for making the saddle-point configurations described herein is the observation that Wilson loops centered on the action peak for a cooled/extremized lump generally have values very close to $-1$, indicating a unit of Z(2) flux. The Wilson loops are roughly of size 10 by 10 for $\beta = 10.0$ and a unit of Z(2) flux is seen for loops in each of the three possible planes. (It is interesting to note that both 't-Hooft[4] and Feynman[5] have speculated about the importance of gauge configurations with Z(2) properties in the vacuum of SU(2) gauge theories in 2+1 dimensions.)

Figure 1 is a sketch of our starting configuration. Relative to a U(1) subgroup of SU(2), it corresponds to a type of monopole-antimonopole pair with the $2\pi$ flux from the monopole channeled into two tubes (each carrying $\pi$ flux) terminating on the anti-monopole. Wilson loops centered on the point $P$ in the middle of the figure have a value of $-1$ for each plane. Only plaquettes inside the closed tube have non-zero action.

Starting from the configuration in Figure 1, we have used extremization to produce a saddle-point configuration with a single peak in the action density. The resulting saddle has the Z(2) properties of the Wilson loops mentioned above. By changing the size of the monopoles and flux tube in the starting configuration, we can get single peak saddle-point configurations with slightly different actions and somewhat different spatial sizes for the peak in the action density. For example, the total action for one of the single-peak configurations (on a $24^3$ lattice) is 1.521 with $\hat{S} = \sum_{\vec{n},i,a,}(\partial S/\partial A(\vec{n})_i^a)^2 = 0.030$. These numbers are extensive quantities, so the amount

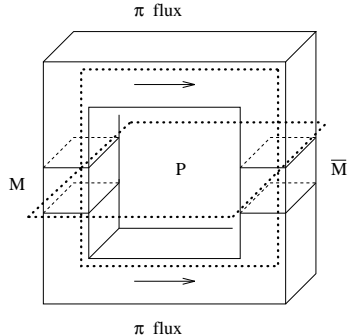

Fig. 1. The starting configuration that produces saddle-point configurations. The monopole and anti-monopole (M and $\bar{\text{M}}$) are defined relative to an arbitrary U(1) subgroup of SU(2). Wilson loops centered on the point P have a value of $-1$. Two such loops are shown by the dotted lines.

by which the field equations are not satisfied at a given point is quite small.

We have also tried extremizing lattices obtained by using two spatially translated copies of the configuration in Figure 1. Figure 2 shows the extremized result from starting with such a configuration. The resulting saddle-point configuration has two isolated peaks in the action density and was extremized until $\hat{S} = 0.0070$. It has an action of 2.948. We have also made $24^3$ lattices with 3 and 4 isolated peaks by using 3 or 4 copies of the configuration in Figure 1. We will refer to these configurations we have been making as a Z(2) saddle gas.

## 3. Maximal Abelian Gauge Monopoles

Because of the interest in determining whether the QCD vacuum is dominated by the condensation of monopoles defined in some particular gauge, we have checked to see what the maximal abelian gauge monopole structure of our saddles is. Table 1 shows the location of the monopoles and antimonopoles for a single-peak configuration. The different entries in the table correspond to choosing a different random gauge before fixing to the maximal abelian gauge.

One can clearly see that the monopole-antimonopole pairs occur near the peak in the action density, which is at $(x, y, z) = (12, 12, 12)$. However, their location depends on which different Gribov copy of the maximal abelian gauge results. In addition, there are some Gribov copies with no monopole-antimonopole pairs. Even though we have not demonstrated that our saddle is important in the QCD vacuum, it gives an example of the difficulty of attributing fundamental significance to monopoles only defined through a gauge fixing procedure.

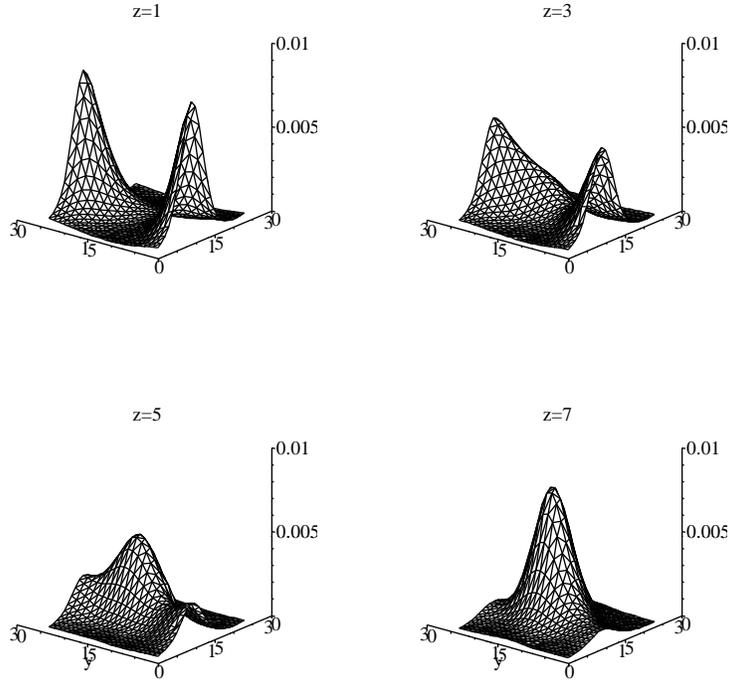

Fig. 2. The action density for 4 $z$-slices of a two-peak saddle on a $24^3$ lattice.

| Random gauge label | Gauge-fixing sweeps | Monopole position | Anti-monopole position | Abelian parameter |
|---|---|---|---|---|
| 1 | 200 | (10,11,10) | (14,9,14) | 0.997804 |
|   | 400 | ( 9,11,10) | (14,11,13) | 0.998064 |
|   | 800 | ( 9,11,10) | (14,11,13) | 0.998076 |
| 2 | 200 | (12,13,9) | (11,8,14) | 0.997844 |
|   | 400 | (11,13,9) | (12,8,14) | 0.997949 |
|   | 800 | (11,13,9) | (12,9,14) | 0.997962 |
| 3 | 200 | none | none | 0.997368 |
|   | 400 | none | none | 0.997384 |
|   | 800 | none | none | 0.997387 |

Table 1. The monopole content (in maximal abelian gauge) of a single-peak saddle-point configuration. Three different random gauges were used as starting points for the gauge fixing to maximal abelian gauge. The right-most column is the value of the quantity being maximized during the gauge fixing; it would be 1 for a configuration with all links in a U(1) subgroup of SU(2). One can see clearly that the monopole content depends on the particular Gribov copy chosen.

## 4. String Tension

Creutz ratios on individual Z(2) saddle-gas configurations are shown in Figure 3, along with the Monte Carlo result and the result for cooled and extremized Monte Carlo lattices. For the two Z(2) saddle-gas configurations shown, one has a continuously rising Creutz ratio, as was seen in extremized Monte Carlo generated lattices. The other has a Creutz ratio which rises and then falls off at large distances.

We are in the process of determining whether averaging over the saddle-point configurations, with the correct weighting, will give a constant Creutz ratio at large distances, indicative of a string tension. We are also exploring whether there are other features of the cooled lattices that we have not included in the saddle-point configurations we are studying currently.

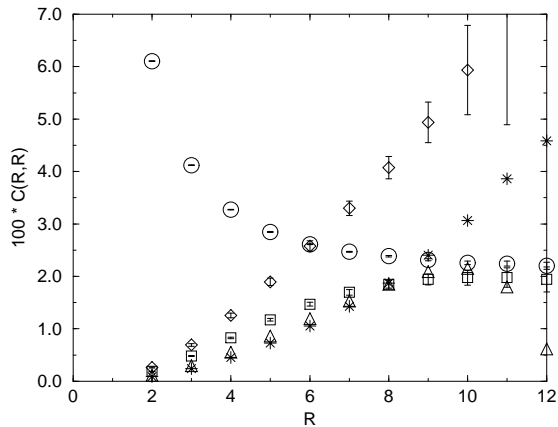

Fig. 3. The Creutz ratios as a function of $R$ for a Monte Carlo calculation at $\beta = 10.0$ ($\circ$), cooled Monte Carlo generated lattices ($\square$), extremized Monte Carlo generated lattices ($\diamond$), and two different 3-peak saddle-point lattices ($\triangle$ and $*$).